\documentclass[12pt]{article}
 \usepackage{epsfig}
 \def\be{\begin{equation}}
 \def\ee{\end{equation}}
 \def\bea{\begin{eqnarray}}
 \def\eea{\end{eqnarray}}
 \usepackage{graphicx}

 \catcode`\@=11
 \def\lsim{\mathrel{\mathpalette\@versim<}}
 \def\gsim{\mathrel{\mathpalette\@versim>}}
 \def\@versim#1#2{\vcenter{\offinterlineskip
 \ialign{$\m@th#1\hfil##\hfil$\crcr#2\crcr\sim\crcr } }}
 \catcode`\@=12

 \catcode`@=12
\begin{document}
 \thispagestyle{empty}
 \begin{flushright}
 UCRHEP-T601\\
 Oct 2020\
 \end{flushright}
 \vspace{0.6in}
 \begin{center}
 {\LARGE \bf Dileptonic Scalar Dark Matter\\
and Exotic Leptons\\}
 \vspace{1.5in}
 {\bf Ernest Ma\\}
 \vspace{0.1in}
{\sl Department of Physics and Astronomy,\\ 
University of California, Riverside, California 92521, USA\\}
\end{center}
 \vspace{1.2in}

\begin{abstract}\
A simple scenario of long-lived dark matter is presented.  Assuming lepton 
number ($L$) conservation with Dirac neutrinos, the neutral component of an 
$SU(2)_L$ scalar triplet with $L=2$ is a suitable candidate, which decays to 
two neutrinos.  Exotic leptons with $L=-1$ may also play a role, and act as 
anchors for small seesaw Dirac neutrino masses.
\end{abstract}

\newpage
\baselineskip 24pt
\noindent \underline{\it Introduction}~:~
Dark matter is a current topic of pervasive research.  For recent reviews, 
see for example Refs.\cite{bh18,l16,y17}.  The conventional wisdom is 
that dark matter is a single particle which was in thermal equilibrium 
with matter in the early Universe and froze out as the temperature of 
the Universe fell below its mass, typically of order 100 GeV to a few TeV.  
Its relic abundance is now precisely known~\cite{planck18}, i.e. 
$\Omega_c h^2 = 0.120 \pm 0.001$.  A dark symmetry is also usually 
assumed to ensure that the lightest dark particle is stable. 

The simplest model of dark matter is that of a stable real singlet scalar 
$S$ which is distinguished from matter by a $Z_2$ symmetry under which only 
$S$ is odd.  For a recent comprehensive study, see Ref.\cite{gambit17}.  The 
scalar potential simply consists of $S$ and the Higgs doublet 
$\Phi = (\phi^+,\phi^0)$ of the Standard Model (SM) of quarks and 
leptons. Assuming that small Majorana neutrino masses are obtained through the 
canonical seesaw mechanism with three heavy singlet right-handed neutrinos 
$N_R$, the terms $S N_R N_R$ are forbidden by the imposed $Z_2$.  There is 
however a better way to understand this dark $Z_2^D$ parity.  It is simply 
derivable~\cite{m15} from lepton parity $Z_2^L = (-1)^L$ as 
$Z_2^D=Z_2^L \times (-1)^{2j}$, where $j$ is the spin of the particle. 
This works by assigning $S$ as odd under $Z_2^L$, i.e. same as the 
leptons.  In other words, $S$ is a leptonic scalar.  Under $Z_2^D$, 
$S$ is then odd, whereas all the leptons (being fermions) are even. 
This simple connection applies to many other scenarios of 
Majorana neutrinos with dark matter, including the original 2006 scotogenic 
one-loop radiative model~\cite{m06}.  In the context of grand unification, 
this leptonic marker is most easily understood as $Q_\chi$ from the 
decomposition $SO(10) \to SU(5) \times U(1)_\chi$~\cite{m18}.

In the continuing absence of experimental evidence for neutrinoless double 
beta decay~\cite{appec19}, there has been a recent surge of theoretical 
interest in Dirac neutrinos, with the recent identification of lepton number 
with a corresponding dark symmetry~\cite{m20}, i.e. a generalization of 
dark parity $Z_2^D$ from lepton parity $Z_2^L$ to $Z_N^D$ from $Z_N^L$ with 
the factor $\omega^{-2j}$ where $\omega^N=1$, as well as $U(1)_D$ from 
$U(1)_L$ with $D = L-2j$.  

More generally, naturally small Dirac neutrino masses~\cite{mp17} may be 
obtained in analogy to Majorana neutrino masses~\cite{m98}.  In any such  
framework, dark matter may also have something to do with lepton number. 
In this paper, a simple scenario is presented by the addition of an 
$SU(2)_L$ scalar triplet $\rho = (\rho^+,\rho^0,\rho^-)$ with $L=2$.

\noindent \underline{\it Dileptonic scalar $SU(2)_L$ triplet}~:~
Consider first $\rho$ with $L=0$.  This would have a trilinear interaction 
with the SM Higgs doublet $\Phi$ through 
$\rho (\Phi^\dagger \Phi)_3$, where
\begin{equation}
(\Phi^\dagger \Phi)_3 = [\bar{\phi}^0 \phi^+, (\phi^- \phi^+ - \bar{\phi}^0 
\phi^0)/\sqrt{2}, \phi^- \phi^0].
\end{equation}
This means that $\rho^0$ will develop a nonzero vacuum expectation value 
and contributes to the $W^\pm$ mass.  It is clearly not a candidate for 
dark matter.

Consider next $\rho$ with $L=1$.  This means that $\rho^-$ is not the 
complex conjugate of $\rho^+$ and that $\rho^0$ is complex, not real. 
It also means that it does not couple to $(\Phi^\dagger \Phi)_3$ or 
any linear combination of two leptons which must have $L=0$ or $L=2$.  
In other words, the designation of $L=1$ is arbitrary for $\rho$. 
This symmetry is in fact new and has nothing to with $L$.

Things are different if $\rho$ has $L=2$.  It may now couple to two leptons 
through the following three dimension-six operators:
\begin{eqnarray}
{\cal L}_1 &=& \Lambda_1^{-2} \rho^\dagger (\Phi \Phi)_3 (L L)_3, \\  
{\cal L}_2 &=& \Lambda_2^{-2} \rho^\dagger (\Phi^\dagger \Phi)_3 \nu_R 
\nu_R, \\  
{\cal L}_3 &=& \Lambda_3^{-2} \partial^\mu(\rho^\dagger \Phi)_2 
\overline{\nu_R^c} \gamma_\mu L,
\end{eqnarray}
where $L = (\nu,e)_L$, 
\begin{eqnarray}
&& (\Phi \Phi)_3 = (\phi^+\phi^+, \sqrt{2}\phi^+ \phi^0, \phi^0 \phi^0), \\ 
&& (L L)_3 = [\nu_L \nu_L, (\nu_L e_L + e_L \nu_L)/\sqrt{2}, e_L e_L], \\ 
&& (\rho^\dagger \Phi)_2 = \left[-\sqrt{2 \over 3}\overline{\rho^-} \phi^0 - 
\sqrt{1 \over 3} \overline{\rho^0} \phi^+, \sqrt{1 \over 3} 
\overline{\rho^0} \phi^0 + \sqrt{2 \over 3} \overline{\rho^+} \phi^+ \right]. 
\end{eqnarray}
The structure of ${\cal L}_3$ is such that it is suppressed by lepton masses 
compared to $v = \langle \phi^0 \rangle$, so it may safely be ignored. 
Now the only two-body decay of $\rho^0$ is to two neutrinos, 
as in an earlier proposal with $Z_3$ lepton symmetry~\cite{mpsz15}. 
In ${\cal L}_1$, $\rho \to \nu_L \nu_L$ has the amplitude 
$-\sqrt{2}v^2/\Lambda_1^2$. In ${\cal L}_2$, $\rho \to \nu_R \nu_R$ has the 
amplitude $-\sqrt{2}v^2/\Lambda_2^2$.  The resulting decay rate is then
\begin{equation}
\Gamma_{1,2} = {m_\rho v^4 \over 16 \pi \Lambda_{1,2}^4},
\end{equation}
assuming the dominant decay to only one Dirac neutrino.  Since $\rho^0$ has 
$L=2$ and lepton number is conserved, other two-body decay modes such as 
$e^+e^-$ and $\gamma \gamma$ are strictly forbidden.  However, three-body 
modes such Higgs + $\nu \nu$ and $W$ + $\nu e$ are possible even though 
they are subdominant.  They would then disrupt the Cosmic Microwave 
Background (CMB) and push the limit of the dark matter's lifetime to 
beyond $10^{25}$ seconds~\cite{sw17}.  As a result,
\begin{equation}
{m_\rho \over 2~{\rm TeV}} < \left( {\Lambda_{1,2} \over 8.63 \times 
10^{14}~{\rm GeV}} \right)^4.
\end{equation}

The components of the scalar triplet $\rho$ are split by their one-loop 
gauge interactions, with the neutral component lower in mass than the 
charged ones~\cite{s95,cfs06} by about 166 MeV.  Hence the latter would 
decay to the former by emitting a virtual $W^\pm$ boson which converts to 
$\pi^\pm$ or $e^\pm,\mu^\pm$ and neutrinos.

As for relic abundance, $\rho$ behaves analogously to fermion triplet  
dark matter~\cite{ms09} as well as the wino in supersymmetry~\cite{b16}. 
Their annihilation to electroweak gauge bosons has the correct cross 
section for $m_\rho$ in a range near 2 TeV.

Since $\rho^0$ does not couple to $Z$ at tree level, it may only interact 
with quarks in one loop through $W^\pm$ exchange, or through the SM Higgs 
boson directly.  These interactions allow it to be detected in 
underground experiments.  With $m_\rho \sim 2$ TeV, a possible discovery 
is on the horizon~\cite{cfs06} of current experiments.

\noindent \underline{\it Ultraviolet completions}~:~
To obtain ${\cal L}_{1,2}$ in a renormalizable theory, new heavy particles 
are required.  Consider first a scalar triplet $\xi = (\xi^{++},\xi^+,\xi^0)$ 
with $L=-2$.  This enables the well-known interaction
\begin{equation}
\xi^0 \nu_L \nu_L - \xi^+(\nu_L e_L + e_L \nu_L)/\sqrt{2} + \xi^{++} e_L e_L.
\end{equation}
The usual next step is to allow $L$ to be broken spontaneously~\cite{gr81} 
or explicitly by the soft term~\cite{sv80,ms98}
\begin{equation}
\xi^0 \bar{\phi}^0 \bar{\phi}^0 - \sqrt{2} \xi^+ \phi^- \bar{\phi}^0 + 
\xi^{++} \phi^- \phi^-,
\end{equation}
in which case the well-known Type II seesaw~\cite{m98} is realized for 
obtaining small Majorana neutrino masses.  Here $L$ is strictly conserved, 
so (10) is allowed but not (11).  With $\rho$ and $\xi$, the triple product
$\rho^\dagger \times (\Phi \Phi)_3 \times \xi^\dagger$ is allowed, 
thereby enabling ${\cal L}_1$ as shown in Fig.~1.  After integrating 
out $\xi$, the dimension-six operator ${\cal L}_1$ is obtained. 
\begin{figure}[htb]
\vspace*{-5cm}
\hspace*{-3cm}
\includegraphics[scale=1.0]{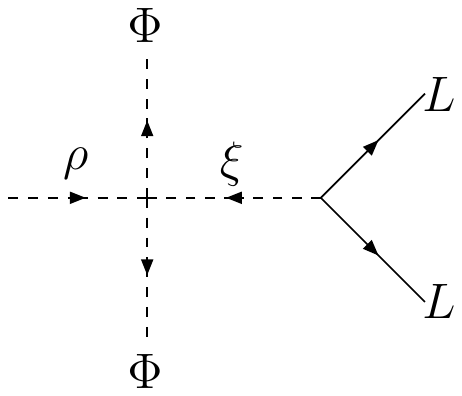}
\vspace*{-21.5cm}
\caption{Dimension-six operator for $\rho^0 \to \nu_L \nu_L$.}
\end{figure}

Consider next a heavy neutral scalar singlet $\zeta$ with $L=-2$, then 
$\zeta \nu_R \nu_R$ is allowed, as well as the quartic scalar interaction 
$\rho^\dagger (\Phi^\dagger \Phi)_3 \zeta^\dagger$.  Together they generate 
Fig.~2, resulting in ${\cal L}_2$.  
\begin{figure}[htb]
\vspace*{-5cm}
\hspace*{-3cm}
\includegraphics[scale=1.0]{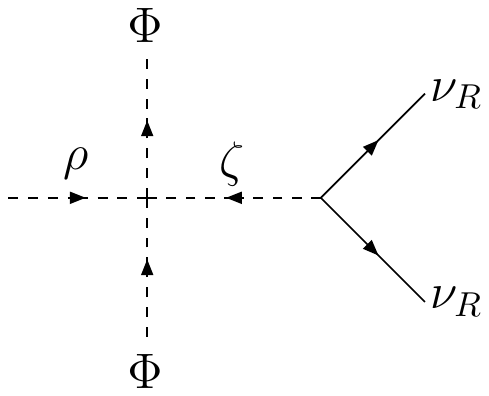}
\vspace*{-21.5cm}
\caption{Dimension-six operator for $\rho^0 \to \nu_R \nu_R$.}
\end{figure}

Note that in a previous proposal~\cite{m18-1}, $\zeta$ is assumed to be 
very light and acts as the dilepton mediator for self-interacting leptonic 
dark matter.  In that case, it must decay very quickly to avoid disrupting 
the standard cosmological scenario of the early Universe.  In the 
singlet-triplet Majoron model with soft breaking of $L$, the 
pseudo-Majoron~\cite{mm17} decays instead to $\nu_L \nu_L$.

\noindent \underline{\it Exotic leptons}~:~
With the new insight of having particles with nonzero lepton numbers, 
an interesting addition to the SM would be the following vectorlike 
doublet $E_{L,R}=(E^0,E^-)_{L,R}$ and singlet $N_{L,R}$ with $L=-1$.  They 
have invariant mass terms $(\overline{E^0_L} E^0_R + \overline{E^-_L} E^-_R)$ 
and $\bar{N}_L N_R$, as well as terms mixing $E$ and $N$ through $\Phi$, 
and those linking them to the SM leptons 
$L=(\nu,e)_L, e_R, \nu_R$ with $L=1$, because $\bar{N}_L$ and $\nu_R$ 
transform identically.  The two sectors mix, through the terms
\begin{equation}
N_R \nu_R, ~~~ (\nu_L \phi^0 - e_L \phi^+) N_L, ~~~ (E^0_R \phi^0 - E^-_R 
\phi^+) \nu_R. 
\end{equation}
This mixing is of course assumed to be very small.  The neutral Dirac 
fermion matrix linking $(\bar{\nu}_L, N_R, E^0_R)$ to 
$(\nu_R, \bar{N}_L, \overline{E_L^0})$ is of the form
\begin{equation}
{\cal M} = \pmatrix{m_\nu & m_{\nu N} & 0 \cr m_{N \nu} & M_N & m_{NE} \cr 
m_{E \nu} & m_{EN} & M_E},
\end{equation}
where $m_{N \nu}$, $m_{E \nu}$, and $m_{\nu N}$ mix the $L=\pm 1$ sectors.

There is however another important coupling, i.e. 
$\rho^\dagger (\bar{E}_R L)_3$, where
\begin{equation}
(\bar{E}_R L)_3 = [\overline{E^-_R} \nu_L, (\overline{E^0_R} \nu_L - 
\overline{E^-_R} e_L)/\sqrt{2}, \overline{E^0_R} e_L].
\end{equation}
Using this, the dimension-six operators ${\cal L}_{1,3}$ are derived 
as shown in Figs.~3 and 4.  
\begin{figure}[htb]
\vspace*{-5cm}
\hspace*{-3cm}
\includegraphics[scale=1.0]{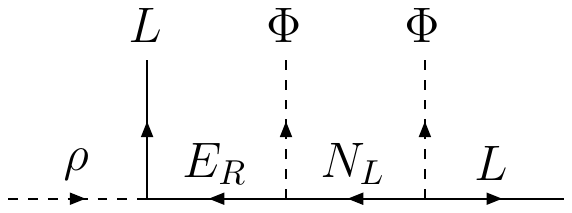}
\vspace*{-21.5cm}
\caption{Dimension-six operator ${\cal L}_1$ using exotic 
leptons.}
\end{figure}
\begin{figure}[htb]
\vspace*{-5cm}
\hspace*{-3cm}
\includegraphics[scale=1.0]{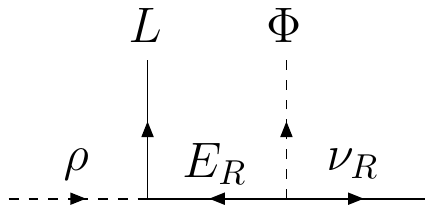}
\vspace*{-21.5cm}
\caption{Dimension-six operator ${\cal L}_3$ using exotic 
leptons.}
\end{figure}
Whereas it appears that only $E_R$ and $N_L$ are needed for these 
operators, the presence of $E_L$ and $N_R$ allows them to have 
heavy invariant masses greater than $m_\rho$ for the validity of 
the interpretation of ${\cal L}_{1,3}$ as effective operators.

\noindent \underline{\it Seesaw Dirac neutrino masses}~:~
The presence of $N_{L,R}$ with $L=-1$ is also useful for obtaining small 
seesaw Dirac neutrino masses.  A $Z_2$ symmetry is imposed~\cite{mp17} 
under which $\nu_R$ (with $L=1$) is odd and all other fields even. 
All dimension-four terms must respect this $Z_2$.  This means that 
$m_{\nu}$ and $m_{E \nu}$ in Eq.~(13) are zero.  However, $Z_2$ is 
allowed to be broken softly, hence $m_{N \nu}$ is nonzero.  Note that 
$L$ is still strictly conserved.  Now ${\cal M}$ of Eq.~(13) becomes 
\begin{equation}
{\cal M} = \pmatrix{0 & m_{\nu N} & 0 \cr m_{N \nu} & M_N & m_{NE} \cr 
0 & m_{EN} & M_E},
\end{equation}
which induces a small Dirac neutrino mass $m_{N \nu} m_{\nu N}/M_N$, 
assuming negligible mixing of $N$ with $E$.

\noindent \underline{\it Conclusion}~:~
If neutrinos are Dirac fermions with strictly conserved lepton number $L$, 
the existence of dark matter may be related to $L$.  A simple 
scenario is presented where an electroweak scalar triplet 
$(\rho^+,\rho^0,\rho^-)$ with $L=2$ is a possible candidate.  The lightest 
component, i.e. $\rho^0$, decays to two neutrinos with a lifetime 
exceeding that of the Universe.  A possible connection to exotic leptons 
with $L=-1$ is also discussed, with the natural appearance of small 
seesaw Dirac neutrino masses..

\noindent \underline{\it Acknowledgement}~:~
I thank Rabi Mohapatra for an important discussion.  This work was supported 
in part by the U.~S.~Department of Energy Grant No. DE-SC0008541.

\bibliographystyle{unsrt}

\end{document}